\documentclass[final]{tlp}

\usepackage{fixtitle}

\usepackage{amssymb}
\usepackage{amsmath}
\usepackage{verbatim}
\usepackage{url}
\usepackage{hhline}
\usepackage{array}
\usepackage{dcolumn}

\newcolumntype{.}{D{.}{.}{-1}}

\newcommand{\defrel}[1]{\mathrel{\buildrel \mathrm{def} \over {#1}}}
\newcommand{\defeq}{\defrel{=}}

\providecommand*{\Nset}{\mathbb{N}}            

\newcommand*{\ts}{\textrm{term-size}}

\newcommand*{\fund}[3]{\mathord{#1}\colon#2\rightarrow#3}

\newcommand*{\Cplusplus}{{C\nolinebreak[4]\hspace{-.05em}\raisebox{.4ex}{\tiny\bf ++}}}

\newcommand*{\cB}{\ensuremath{\mathcal{B}}}
\newcommand*{\cH}{\ensuremath{\mathcal{H}}}
\newcommand*{\cN}{\ensuremath{\mathcal{N}}}

\newcommand*{\pre}{\mathrm{pre}}
\newcommand*{\post}{\mathrm{post}}

\newenvironment{Pro}[0]
{\begin{minipage}[t]{40ex}%
        \begin{tabbing}%
\hspace{3ex}\=\hspace{3ex}\=\hspace{3ex}\=\hspace{3ex}\=%
\hspace{6ex}\=\hspace{6ex}\=\hspace{6ex}\=\hspace{6ex}\=%
\hspace{6ex}\=\hspace{6ex}\=\hspace{6ex}\=\hspace{3ex}\kill}%
{\end{tabbing}%
\end{minipage}%
}

\begin{document}

\newcommand\othrpearl{Technical Note}
\title{cTI: A Constraint-Based \\ Termination Inference Tool \\ for ISO-Prolog}
\author[F. Mesnard and R. Bagnara]{
       {FRED MESNARD}
       \affiliation
       Iremia,
       Universit\'e de La R\'eunion,
       Saint Denis, France \\
       \email{fred@univ-reunion.fr}
       \and
       {ROBERTO BAGNARA}
       \affiliation
       Department of Mathematics,
       University of Parma,
       Italy \\
       \email{bagnara@cs.unipr.it}
}

\submitted{11 July 2001}
\revised{11 March 2003 and 10 September 2003}
\accepted{12 September 2003}
\pubyear{2003}

\maketitle
\shorttitle{cTI: A Termination Inference Tool for Prolog}

\begin{abstract}
We present cTI, the first system for universal left-termination
inference of logic programs.  Termination inference generalizes
termination analysis and checking.  Traditionally, a termination
analyzer tries to prove that a given class of queries terminates.
This class must be provided to the system, for instance by means of
user annotations.  Moreover, the analysis must be redone every time
the class of queries of interest is updated.  Termination inference,
in contrast, requires neither user annotations nor recomputation.  In
this approach, terminating classes for all predicates are inferred at
once.  We describe the architecture of cTI and report an extensive
experimental evaluation of the system covering many classical examples
from the logic programming termination literature and several Prolog
programs of respectable size and complexity.
\end{abstract}

\section{Introduction}
\label{Intro}

Termination is a crucial aspect of program verification.
It is of particular importance for logic programs \cite{Lloyd87,Apt97},
since there are no \textit{a priori} syntactic restrictions to queries
and, as a matter of fact, most predicates programmers tend to write
do not terminate for their most general queries.
In the last fifteen years, termination has been the subject of several
research works in the field of logic programming (see, for instance,
\cite{FrancezGKP85,AptP90,Ruggieri99th}).
In contrast to what happens for other programming paradigms,
there are two notions of termination for logic programs \cite{VasakP86}:
\emph{existential} and \emph{universal} termination.
To illustrate them, assume we are using a standard Prolog engine.
Existential termination of a query means that either the computation
finitely fails or it produces \emph{one} solution in finite time.
This does not exclude the possibility that the engine, when asked
for further solutions, will loop.
On the other hand, universal termination means that the computation
yields all solutions and eventually fails in finite time (if we
repeatedly ask for further solutions).

Although the concept of existential termination plays an important role
in connection with \emph{normal} logic programs, it has severe drawbacks
that make it not appropriate in other contexts:
existential termination is not \emph{instantiation-closed} (i.e., a goal may
existentially terminate, yet some of its instances may not terminate),
hence it is not \emph{and-compositional} (i.e., two goals may existentially
terminate while their conjunction does not); finally, existential
termination depends on the textual order of clauses in the program.
Universal termination is a stronger and much more robust concept:
it implies existential termination and it is both and-compositional
and instantiation-closed.

Existential termination has been the subject of only a few works
\cite{VasakP86,LeviS95,Marchiori96} whereas most research focused on
universal termination.
There are two main directions (see \cite{DeSchreyeD94} for a
survey): characterizing termination \cite{AptP90,AptP93,Ruggieri99th}
and finding weaker but decidable sufficient conditions
that lead to actual algorithms,
e.g., \cite{UllmanVG88,Plumer90lncs,Verschaetse92th}.
Even though our research belongs to both streams, in this paper we
focus on an intuitive presentation of the implementation of our
approach.
A companion paper presents a complete formalization of our work in the
theoretical setting of acceptability for constraint logic programs
\cite{MesnardR03}, where we refine a necessary and sufficient condition
for termination to the sufficient condition implemented in cTI.

Our main contribution compared to other automated
termination analyzers
\cite{LindenstraussS97,Decorte97th,SpeirsSS97,CodishT99} is that our tool
\emph{infers} sufficient universal termination conditions from the
text of any Prolog program,
adopting a bottom-up approach to termination.
An important feature of this approach first presented in
\cite{Mesnard96} is that there is no need to define in advance a class
of queries of interest.
(If required, these classes can be provided after the analysis has finished
in order to specialize the obtained results.)
Our system,
called \emph{cTI} from \emph{\underline{c}onstraint-based
\underline{T}ermination \underline{I}nference},
is written in SICStus Prolog.
A preliminary account of the  work
described in this paper appeared in \cite{MesnardN01},
where we showed that numeric computations took
most of the execution times.
Now cTI relies on the specialized Parma Polyhedra Library
\cite{BagnaraRZH02}, a modern \Cplusplus{} library for the manipulation
of convex polyhedra that significantly speeds up the analysis.
Moreover, cTI has been extended so that it can analyze any ISO-Prolog program
\cite{ISO-Prolog-part-1,Deransart96}.
The only correctness requirement we currently impose on programs is that
they must not create infinite rational terms.
Hence we assume execution with occurs-check or, equivalently,
NSTO programs (i.e., programs that are \emph{Not Subject to Occur-Check}
\cite{DeransartFT91} and thus are safely executed with any standard conforming
system).
We point out that simple, sufficient syntactic methods for ensuring
occurs-check freedom are presented in \cite{AptP94} while
\cite{Sondergaard86,CrnogoracKS96} describe abstract-interpretation based
solutions.
Recently, \emph{finite-tree analysis} \cite{BagnaraGHZ01,BagnaraZGH01}
has been proposed to confine infinite rational terms in programs
that are not occurs-check free.
Both the approach described in \cite{MesnardR03} and the cTI system
can be extended, with the help of finite-tree analysis,
to deal also with such programs.

Throughout the paper we assume a basic knowledge of
logic programming (see, e.g., \cite{Apt97}),
constraint logic programming (see, e.g., \cite{MarriottS98}),
abstract interpretation (see, e.g., \cite{CousotC92lp}), and
propositional $\mu$-calculus (see, e.g., \cite{ClarkeGP00}).
In Section~\ref{sec:cti-overview} we present cTI informally with an
example analysis.
How to use cTI is described in Section~\ref{sec:using-cti}.
An experimental evaluation of the system is the subject of
Section~\ref{sec:experimental-evaluation}.
Related work is discussed in Section~\ref{sec:related-work}
while Section~\ref{sec:conclusion} concludes.

\section{An Overview of cTI}
\label{sec:cti-overview}

Our aim is to compute classes of queries for which universal
left termination  is guaranteed. We call such classes
\emph{termination conditions}. More precisely,
let $P$ be a Prolog program and $q$ a predicate symbol of $P$.
A termination condition for $q$ is
a set $\mathrm{TC}_q$ of goals of the form $\leftarrow c,q(\tilde{x})$
where $c$ is a CLP(\cH) constraint such that, for any goal
$G \in \mathrm{TC}_q$,
each derivation of $P$ and
$G$ using the left-to-right selection rule is finite.

Our analyzer uses three main constraint structures:
Herbrand terms for the initial program $P$
(seen as a CLP(\cH) program),
non-negative integers, and booleans
($P$ is abstracted into both a CLP(\cN)
and a CLP(\cB) program).
We illustrate our method to infer termination conditions by means
of an example.
The method consists of six distinct steps,
which will be illustrated on the following definition for
the predicates \texttt{app/3}, \texttt{nrev/2} and \texttt{app3/4}.

\medskip
\noindent
\begin{tabular}{l|l|l}
\begin{Pro}
\verb/app([], X, X)./          \\
\verb/app([E|X], Y, [E|Z]) :- / \\
\>      \verb/app(X, Y, Z)./
\end{Pro}
&
\begin{Pro}
\verb/nrev([], [])./           \\
\verb/nrev([E|X], Y) :- /       \\
\>      \verb/nrev(X, Z),/     \\
\>      \verb/app(Z, [E], Y)./
\end{Pro}
&
\begin{Pro}
\verb/app3(X, Y, Z, U) :- /     \\
\>      \verb/app(X, Y, V),/   \\
\>      \verb/app(V, Z, U)./
\end{Pro}
\end{tabular}

\paragraph{Step~1: From Prolog to CLP(\cN).}
From the Prolog program $P$, a CLP(\cN) program $P^\cN$
is obtained by applying a symbolic norm.
In our example, we use the \emph{term-size} norm,
which is the one cTI applies by default.
All ISO-predefined predicates have been manually pre-analyzed
for this norm.
Notice that, as explained in \cite{MesnardR03},
termination inference for pure Prolog programs can be based
on any linear norm.
The symbolic term-size norm is inductively defined as follows:
\[
  \|t\|_\ts
    \defeq
      \begin{cases}
        1 + \sum_{i=1}^n \|t_i\|_\ts,
            &\text{if $t = f(t_1, \ldots, t_n)$ with $n > 0$;} \\
        0,  &\text{if $t$ is a constant;} \\
        t,  &\text{if $t$ is a variable.}
      \end{cases}
\]
For example, $\|f(0,0)\|_\ts = 1$.
All non-monotonic elements of the program are
approximated by monotone constructs.
For instance, Prolog's unsound negation
\verb/\+ G/ is approximated by \verb/((G, false) ; true)/.
More generally, extra-logical predicates are mapped to their
first-order counterparts so that the termination property is
preserved.
For our running example, we obtain the following CLP(\cN) clauses:

\medskip
\begin{tabular}{l|l|l}
\begin{Pro}
$\mathrm{app}_\cN(0, x, x).$                       \\
$\mathrm{app}_\cN(1+e+x,y,1+e+z) \leftarrow$   \\
\>     $\mathrm{app}_\cN(x, y, z).$
\end{Pro}
&
\begin{Pro}
$\mathrm{nrev}_\cN(0, 0).$                            \\
$\mathrm{nrev}_\cN(1+e+x,y) \leftarrow$          \\
\>      $\mathrm{nrev}_\cN(x, z),$                  \\
\>      $\mathrm{app}_\cN(z, 1+e, y).$
\end{Pro}
&
\begin{Pro}
$\mathrm{app3}_\cN(x, y, z, u) \leftarrow$ \\
\>      $\mathrm{app}_\cN(x, y, v),$       \\
\>      $\mathrm{app}_\cN(v, z, u).$
\end{Pro}
\end{tabular}

\paragraph{Step~2: Computing a numeric model.}
A model of the CLP(\cN) program is now computed.
For each predicate $p$, the model describes, with a finite
conjunction of linear equalities and inequalities
denoted by $\post_p^\cN$,
the linear inter-argument relations that hold for every solution of $p$.
In our example we obtain the following
model:
\begin{align*}
  \post^\cN_\mathrm{app}(x,y,z)    &\iff x + y = z,\\
  \post^\cN_\mathrm{nrev}(x,y)     &\iff x = y, \\
  \post^\cN_\mathrm{app3}(x,y,z,u) &\iff x + y + z = u.
\end{align*}
The actual computation is performed on the set of nonnegative,
infinite precision rational numbers, using a  fixpoint calculator
based on PPL, the Parma Polyhedra Library \cite{BagnaraRZH02},
and the standard widening \cite{CousotH78,Halbwachs79th}.
In our example the least model is found.
In general, however, only a less precise model can be determined.

\paragraph{Step~3: Computing a numeric level mapping.}
The information provided by the numerical model is crucial
to compute a \emph{level mapping} $|\cdot|^\cN$.
Let $p$ be an $n$-ary predicate symbol in the CLP(\cN) program.
The level mapping associates to $p$ a function $\fund{f_p}{\Nset^n}{\Nset}$
that is guaranteed to decrease when going from the head of the clause
to each recursive call(s), if any, for each clause defining $p$.
For example, a level mapping $|\cdot|^\cN$ such that
$|\mathrm{nrev}_\cN(x,y)|^\cN = x$ intuitively means:
for each ground instance\footnote{That is, where natural numbers
have replaced variable symbols.} of each recursive clause
defining $\mathrm{nrev}_\cN$,
the first argument decreases when going from the head of the clause
to the recursive call
(since $1+e+x > x$ for each $e,x \in \Nset$).
Since no clause defining $\mathrm{app3}_\cN$ is recursive,
the level mapping can be defined so that $|\mathrm{app3}(x,y,z,t)|^\cN=0$.
The level mapping computed for our example is defined by:
\begin{align*}
  |\mathrm{app}(x,y,z)|^\cN    &= \min(x,z), \\
  |\mathrm{nrev}(x,y)|^\cN     &= x, \\
  |\mathrm{app3}(x,y,z,u)|^\cN &= 0.
\end{align*}
This is obtained by means of an improvement of the technique
by K.~Sohn and A.~Van~Gelder for the automatic generation
of linear level mappings.
Their algorithm, which is based on linear programming,
is complete in the sense that it will always provide
a linear level mapping if one exists \cite{SohnVG91}.
Our extension, which is described in \cite{MesnardN01}, consists in
first computing a constraint over the coefficients of a generic linear
level mapping (step~3a).
Then  we generate a concrete level mapping (step~3b).
Notice that for a multi-directional predicate (such as \texttt{app/3})
we may get multiple linear level mappings.
These are combined, with the $\min$ operator, into
one non-linear level mapping.

In contrast with the well-known standard framework of acceptability,
the decrease of the level mapping has to be shown
only for predicates belonging to the same
strongly connected
component (SCC) of the call graph.
Step~5 below will ensure that the other calls to predicates from lower
SCC's do left terminate.
The advantage of this approach is twofold:
first, the computation of a level mapping, being SCC-based, is modular.
Secondly, the expressive power of \emph{linear} level mappings with
respect to termination is much higher than in the acceptability case.

\paragraph{Step~4: From CLP(\cN) to CLP(\cB).}
From the CLP(\cN) program $P^\cN$
a CLP(\cB) program,  $P^\cB$, is obtained by mapping
each natural number to $1$ (true),
each variable symbol to itself,
and addition to logical conjunction.

\medskip
\begin{tabular}{l|l|l}
\begin{Pro}
$\mathrm{app_\cB}(1, x, x).$                                               \\
$\mathrm{app}_\cB(1 \land e \land x, y, 1 \land e \land z) \leftarrow$ \\
\>      $\mathrm{app}_\cB(x, y, z).$
\end{Pro}
&
\begin{Pro}
$\mathrm{nrev}_\cB(1, 1).$                            \\
$\mathrm{nrev}_\cB(1 \land e \land x,y) \leftarrow$          \\
\>      $\mathrm{nrev}_\cB(x, z),$                  \\
\>      $\mathrm{app}_\cB(z, 1 \land e, y).$
\end{Pro}\ &
\begin{Pro}
$\mathrm{app3}_\cB(x, y, z, u) \leftarrow$ \\
\>      $\mathrm{app}_\cB(x, y, v),$       \\
\>      $\mathrm{app}_\cB(v, z, u).$
\end{Pro}
\end{tabular}
\smallskip

\noindent
The purpose of $P^\cB$ is the one of capturing boundedness dependencies
within $P^\cN$ or, equivalently, rigidity dependencies
within the original program.\footnote{A term $t$ is \emph{rigid}
with respect to
a symbolic norm $\|\cdot\|$ if and only if its measure is invariant
by instantiation, i.e., $\|t\| = \|t\theta\|$ for any substitution $\theta$.}
A model for $P^\cB$ is then computed and a boolean level mapping
$|\cdot|^\cB$ is obtained from the numerical level mapping computed
in Step~3.  In order to do that, the translation scheme outlined
above is augmented with the association of the logical disjunction
$x \lor y$ to $\min(x, y)$: this means that $\min(x, y)$ is a bounded
quantity if $x$ or $y$ or both are bounded.
Here is what we obtain for the example program:
\begin{align*}
  \post^\cB_\mathrm{app}(x,y,z)    &\iff (x \land y) \leftrightarrow z,
&
  |\mathrm{app}(x,y,z)|^\cB    &= x \lor z, \\
  \post^\cB_\mathrm{nrev}(x,y)     &\iff x \leftrightarrow y,
&
  |\mathrm{nrev}(x,y)|^\cB     &= x, \\
  \post^\cB_\mathrm{app3}(x,y,z,u) &\iff (x \land y \land z) \leftrightarrow u,
&
  |\mathrm{app3}(x,y,z,u)|^\cB &= 1.
\end{align*}
For instance, as we use the term-size norm, this  model tells us that for
any computed answer $\theta$ to a call $\texttt{nrev}(x,y)$,
$x\theta$ is ground if and only if $y\theta$ is ground.

\paragraph{Step~5: Computing boolean termination conditions.}
The information obtained from $P^\cB$ for each program point
is combined with the level mapping by means of the following boolean
$\mu$-calculus formul\ae, whose solution gives the desired boolean termination
conditions.

\newcommand*{\itc}{\mathrel:}
\newcommand*{\st}{\mathrel.}
\begin{align*}
  \pre_{\mathrm{app}}
    &=
      \nu T \st \lambda (x, y, z) \st \\
    &
      \begin{cases}
        \bigl|\mathrm{app}(x, y, z)\bigr|^\cB \\
        \forall e, x', z' \itc
          \Bigl(
            \bigl(x \leftrightarrow (1 \land e \land x')\bigr)
            \land
            \bigl(z \leftrightarrow (1 \land e \land z')\bigr)
          \Bigr) \rightarrow T(x', y, z')
      \end{cases} \\
  \pre_{\mathrm{nrev}}
    &=
      \nu T \st \lambda (x,y) \st \\
    &
      \begin{cases}
        |\mathrm{nrev}(x,y)|^\cB & (1) \\
        \forall e, x', z \itc
          \Bigl(
            \bigl(x \leftrightarrow (1 \land e \land x')\bigr)
          \Bigr) \rightarrow T(x', z) & (2) \\
        \forall e, x', z \itc
          \Bigl(
            \bigl(x \leftrightarrow (1 \land e \land x')\bigr)
            \land
              \post^\cB_{\mathrm{nrev}}(x',z)
          \Bigr)
            \rightarrow \pre_{\mathrm{app}}(z, 1 \land e ,y) \quad\;& (3)
      \end{cases} \\
  \pre_{\mathrm{app3}}
    &=
      \nu T \st \lambda(x, y, z, u) \st \\
    &
      \begin{cases}
        \bigl|\mathrm{app3}(x, y, z, u)\bigr|^\cB \\
          \forall v \itc 1 \rightarrow \pre_{\mathrm{app}}(x, y, v) \\
          \forall v \itc \post^\cB_{\mathrm{app}}(x,y,v)
            \rightarrow \pre_{\mathrm{app}}(v,z,u)
      \end{cases}
\end{align*}
Here is the intuition behind such boolean $\mu$-calculus formul\ae.
Consider the \verb+nrev/2+ predicate.
Its unit clause is taken into account in the computation of
the numeric and the boolean model.
For computing the boolean termination condition $\pre_{\mathrm{nrev}}$,
we consider the clause
\[
  \mathrm{nrev}_\cB(x,y)
    \leftarrow
      [x \leftrightarrow (1 \land e \land x')],
      \mathrm{nrev}_\cB(x', z),
      \mathrm{app}_\cB(z, 1\land e, y).
\]
We are looking for a boolean relation $T(x,y)$
satisfying the following conditions:
\begin{itemize}
\item
for each $(x,y)$ in $T$,
the level mapping has to be bounded, which leads to condition (1) above;
\item
the recursive call to \verb+nrev/2+
has to terminate, hence condition (2);
\item
for any state resulting from the evaluation of the first call,
the subsequent call to \verb+app/3+ has to terminate,
giving condition (3);
\item
finally, we are interested in the weakest solution for $T$,
hence the boolean termination condition is defined as
a greatest fixpoint:
\[
  \pre_{\mathrm{nrev}}
    =
      \nu T \st \lambda (x,y) \st \bigl\{ (1) \land (2) \land (3) \bigr\}.
\]
\end{itemize}
Solving the equations for our example gives:
\begin{align*}
  \pre_{\mathrm{app}}(x,y,z)    &= x \lor z, \\
  \pre_{\mathrm{nrev}}(x,y)     &= x, \\
  \pre_{\mathrm{app3}}(x,y,z,u) &= (x \land y) \lor (x \land u).
\end{align*}

The greatest fixpoint is evaluated with the boolean $\mu$-solver
described in \cite{ColinMR97}, which
computes on the domain $\mathrm{Pos}$ of positive boolean formul\ae
~\cite{ArmstrongMSS98}
and is based on the boolean solver of SICStus Prolog.

\paragraph{Step 6: Back to Prolog.}
In the final step of the analysis, the boolean termination conditions
are lifted to termination conditions with the following interpretation,
where the $c$'s are CLP(\cH) constraints:
\begin{itemize}
\item
each goal `\verb+?-+~$c$,~\verb+app(X,Y,Z).+'
left-terminates if \verb+X+ or \verb+Z+ are ground in $c$;
\item
each goal `\verb+?-+~$c$,~\verb+nrev(X,Y).+' left-terminates
if \verb+X+ is ground in $c$;
\item
each goal `\verb+?-+~$c$,~\verb+app3(X,Y,Z,U).+' left-terminates
if \verb+X+ and \verb+Y+ are ground in $c$
or \verb+X+ and \verb+U+ are ground in $c$.
\end{itemize}

\section{Using cTI}
\label{sec:using-cti}

Once compiled and installed, cTI is invoked with the command
`\verb+cti source+', where the program in `\verb+source+' is assumed
to be an ISO-Prolog program.
The user may then control the behavior of cTI with some options.
We describe the main ones.

\begin{description}
\item[`\texttt{-p file}']
By default, undefined predicates are assumed to fail.
The user may enrich or redefine the set of built-ins recognized by the
system, by specifying `\verb+-p file+' on the command line.
This has the effect of importing the predicates whose numerical model,
boolean model, and termination condition are given in `\verb+file+'.
As predicates imported that way cannot be redefined in the
analyzed program, this scheme provides a way to overcome potential
weaknesses of the analysis.
\item[`\texttt{-t timeout\_in\_ms}']
The analysis steps~2,~3a,~3b,~4,~and~5 described in
Section~\ref{sec:cti-overview} all include potentially expensive
computations.
Because of this, for each such step, the computation concerning each SCC
is subject to a timeout, whose default value is 2 seconds.
The `\verb+-t+' option allows the user to modify this value.
\item[`\texttt{-n N}']
For the computation of the numeric model (step~2), a widening is used
after $n$ iterations of the approximate fixpoint iteration.
The default value for $n$ is 1.
\end{description}

The user may also modify a  program to give specific information for
selected program points.
We illustrate this facility by means of examples;
the precise syntax is given in the cTI's documentation.
One may specify that particular program variables will only be bound
to non-negative integers and that the analyzer should take into account 
some constraints involving them.
For instance, cTI does not detect that the following program terminates:
\begin{verbatim}
p1(N) :- N > 0, M is N-1, p1(M).
p1(N) :- N > 1, A is N>>1, Z is N-A, p1(A), p1(Z).
\end{verbatim}
where the predefined arithmetic functor `\texttt{>>/2}' is the bitwise
arithmetic right shift.
On the other hand, cTI is able to show that \verb+p2(N)+
terminates:
\begin{verbatim}
p2(N) :- cti:{N > 0, M = N-1}, p2(M).
p2(N) :- cti:{N > 1, 2*A =< N, N =< 2*A+1, Z = N-A}, p2(A), p2(Z).
\end{verbatim}
Finally, at any program point, the user can add linear inter-argument
relations or groundness relations that the analyzer will take for
granted.
The system can thus prove the termination of the goal `\verb+?- top.+'
where the predicate \verb+top/0+ is defined by the program given in
Section~\ref{sec:cti-overview} augmented with the following clause,
where the term-size of \verb+L1+ is declared to be less than 10
and \verb+L2+ is declared to be ground:
\begin{verbatim}
top :- cti:{n(L1) < 10},app(L1,Zs,L2),cti:{b(L2)},app(Xs,Ys,L2).
\end{verbatim}
While such programs are no longer  ISO-Prolog programs,
the annotations can be automatically removed so as to obtain
the original programs back.
The assertion language currently used in cTI is only experimental,
and future versions of the system may be based on the language defined in
\cite{HermenegildoPB2000}.

\section{Experimental Evaluation}
\label{sec:experimental-evaluation}

Unless otherwise specified, the experiments we present here
were all conducted with the option (see Section \ref{sec:using-cti})
\verb+-p predef_for_compatibility.pl+,
which ensures that non-ISO built-ins used in the benchmarks
(several of which are written in a non-ISO dialect of Prolog)
are predefined.
This experimental evaluation was done on a GNU/Linux system
with an Intel i686 CPU clocked at 2.4 GHz, 512 Mb of RAM,
running the Linux kernel version 2.4,
SICStus Prolog 3.10.0 (28.3 MLips),
PPL version 0.5, and cTI version 1.0.

\paragraph{Standard programs from the termination literature.}
Table~\ref{table_results_dds} presents timings and results
of cTI on some  standard LP termination benchmarks.
The columns are labeled as follows:
\begin{description}
\item[program:]
the name of the analyzed program (the asterisk near a name means
that we had to use one of the options that allow to tune the
behavior of cTI);
\item[top-level predicate:]
the predicate of interest;
\item[checked:]
the class of queries \emph{checked} by the analyzers
of \cite{DecorteDSV99,LindenstraussS97,SpeirsSS97};
\item[result:]
the best result among those reported in
\cite{DecorteDSV99,LindenstraussS97,SpeirsSS97} (where, of course,
`\emph{yes, the program terminates}' is better than
`\emph{no, don't know}');
\item[inferred:]
the termination condition \emph{inferred} by cTI
(1 means that any call to the predicate terminates, 0 means that cTI
could not find a terminating mode for that predicate);
\item[time:]
the running time, in seconds, for cTI to infer the termination conditions.
\end{description}

For all the examples presented in Table~\ref{table_results_dds},
our analyzer is able
to infer a class of terminating queries at least as large than
the one checked by the analyzers of
\cite{DecorteDSV99,LindenstraussS97,SpeirsSS97}
(although we manually tuned cTI three times).
We point out that TermiLog \cite{LindenstraussS97} and TerminWeb
\cite{CodishT99} are sometimes able to prove termination whereas cTI
is not and \emph{vice versa}.

\begin{table}
\renewcommand{\arraystretch}{1.2}
\centering
\caption{De Schreye's, Apt's, and Pl\"umer's programs.}
\label{table_results_dds}
\footnotesize
\begin{tabular}{||l|l||c|c||c|.||}
\hhline{~~|t:==:t:==:t|}
\multicolumn{2}{c||}{}
  & \multicolumn{2}{c||}{Others}
  & \multicolumn{2}{c||}{cTI} \\
\hhline{|t:==::==::==:|}
program & top-level predicate & checked & result & inferred
 & \multicolumn{1}{c||}{time (s)} \\
\hhline{|:==::==::==:|} 
\tt permute             &$\mathrm{permute}(x,y)$           &$x$       &yes&$x$               &0.03\\
\tt duplicate           &$\mathrm{duplicate}(x,y)$         &$x$       &yes&$x\lor y$         &0.02\\
\tt sum                 &$\mathrm{sum}(x,y,z)$             &$x\land y$&yes&$x\lor y \lor z$  &0.03\\
\tt merge               &$\mathrm{merge}(x,y,z)$           &$x\land y$&yes&$(x\land y)\lor z$&0.03\\
\tt dis-con             &$\mathrm{dis}(x)$                 &$x$       &yes&$x$               &0.03\\
\tt reverse             &$\mathrm{reverse}(x,y,z)$         &$x\land z$&yes&$x$               &0.02\\
\hhline{||--||--||--||} 
\tt append              &$\mathrm{append}(x,y,z)$          &$x\land y$&yes&$x\lor z$         &0.02\\
\tt list                &$\mathrm{list}(x)$                &$x$       &yes&$x$               &0.01\\
\tt fold                &$\mathrm{fold}(x,y,z)$            &$x\land y$&yes&$y$               &0.02\\
\tt lte                 &$\mathrm{goal}$                   &$1$       &yes&$1$               &0.02\\
\tt map                 &$\mathrm{map}(x,y)$               &$x$       &yes&$x \lor y$        &0.02\\
\tt member              &$\mathrm{member}(x,y)$            &$y$       &yes&$y$               &0.01\\
\tt mergesort           &$\mathrm{mergesort}(x,y)$         &$x$       &no &$0$               &0.06\\
\tt mergesort\textsuperscript{*}       &$\mathrm{mergesort}(x,y)$       &$x$       &no &$x$               &0.07\\
\tt mergesort\_ap       &$\mathrm{mergesort\_ap}(x,y,z)$   &$x$       &yes&$z$               &0.11\\
\tt mergesort\_ap\textsuperscript{*}   &$\mathrm{mergesort\_ap}(x,y,z)$ &$x$       &yes&$x\lor z$         &0.11\\
\tt naive\_rev          &$\mathrm{naive\_rev}(x,y)$        &$x$       &yes&$x$               &0.03\\
\tt ordered             &$\mathrm{ordered}(x)$             &$x$       &yes&$x$               &0.01\\
\tt overlap             &$\mathrm{overlap}(x,y)$           &$x\land y$&yes&$x \land y$       &0.01\\
\tt permutation         &$\mathrm{permutation}(x,y)$       &$x$       &yes&$x$               &0.03\\
\tt quicksort           &$\mathrm{quicksort}(x,y)$         &$x$       &yes&$x$               &0.06\\
\tt select              &$\mathrm{select}(x,y,z)$          &$y$       &yes&$y \lor z$        &0.01\\
\tt subset              &$\mathrm{subset}(x,y)$            &$x\land y$&yes&$x \land y$       &0.02\\
\tt sum                 &$\mathrm{sum}(x,y,z)$             &$z$       &yes&$y\lor z$         &0.02\\
\hhline{||--||--||--||} 
\tt pl2.3.1     &$\mathrm{p}(x,y)$          &$x$                     &no &$0$                      &0.01\\
\tt pl3.5.6     &$\mathrm{p}(x)$            &$1$                     &no &$x$                      &0.01\\
\tt pl3.5.6a    &$\mathrm{p}(x)$            &$1$                     &yes&$x$                      &0.01\\
\tt pl4.0.1     &append3(x,y,z,v)&$x\land y\land z $      &yes&$(x\land y)\lor (x\land v)$  &0.02\\
\tt pl4.5.2     &$\mathrm{s}(x,y)$          &$x$                     &no &$0$                      &0.03\\
\tt pl4.5.3a    &$\mathrm{p}(x)$            &$x$                     &no &$0$                      &0.01\\
\tt pl5.2.2    &$\mathrm{turing}(x,y,z,t)$ &$x\land y\land z$       &no &$0$                      &0.11\\
\tt pl7.2.9     &$\mathrm{mult}(x,y,z)$     &$x\land y$              &yes&$x\land y$               &0.02\\
\tt pl7.6.2a    &$\mathrm{reach}(x,y,z)$    &$x\land y\land z$       &no &$0$                      &0.02\\
\tt pl7.6.2b    &$\mathrm{reach}(x,y,z,t)$  &$x\land y\land z\land t$&no &$0$                      &0.03\\
\tt pl7.6.2c    &$\mathrm{reach}(x,y,z,t)$  &$x\land y\land z\land t$&yes&$z\land t$               &0.04\\
\tt pl8.3.1     &$\mathrm{minsort}(x,y)$    &$x$                     &no &$x\land y$               &0.04\\
\tt pl8.3.1a    &$\mathrm{minsort}(x,y)$    &$x$                     &yes&$x$                      &0.04\\
\tt pl8.4.1     &$\mathrm{even}(x)$         &$x$                     &yes&$x$                      &0.02\\
\tt pl8.4.2     &$\mathrm{e}(x,y)$          &$x$                     &yes&$x$                      &0.07\\
\hhline{|b:==:b:==:b:==:b|}
\end{tabular}
\end{table}

\paragraph{Standard programs from the abstract interpretation literature.}
Table~\ref{table_results_middle}
presents timings of cTI using some
standard benchmarks\footnote{These have been collected by N.~Lindenstrauss,
see \url{www.cs.huji.ac.il/~naomil}.}
from the LP program analysis community.
We have chosen eleven middle-sized, well-known logic programs. All
the programs are taken from
\cite{BuenoDLBH94} except {\tt credit} and {\tt plan}.
The first column of Table~\ref{table_results_middle} gives the name of
the analyzed program and the second one gives the number of its clauses
(before any program transformation takes place).
The following six columns indicate the running times
(minimum execution times over ten runs), in seconds,
for computing:
\begin{description}
\item[$M_P^{\cN}$:] a numeric model (step~2);
\item[$C_{\mu}$:] the
constraint over the coefficients of a generic linear
level mapping (step~3a);
\item[$\mu$:]
the concrete level mapping (step~3b);
\item[$M_P^{\cB}$:]
a boolean model (step~4);
\item[$\mathrm{TC}$:]
 the boolean  termination conditions (step~5).
\end{description}
The next column reports the total runtime in seconds
while the last column, labeled `Q\%',
expresses the quality of the analysis,
computed as the ratio of the number of user-defined predicates that have
a non-empty termination condition over the total number of
user-defined predicates
(the result of an analysis presents all the user-defined predicates together
with their corresponding termination conditions).

We note that cTI can prove that
{\tt bid}, {\tt credit}, and {\tt plan} are \emph{left-terminating}:
every ground atom left-terminates.
For any such program $P$, $T_P$ has only one fixpoint
\cite[Theorem~8.13]{Apt97}, which may help proving its partial correctness.
Moreover, as the ground semantics of such a program is decidable,
Prolog is its own decision procedure, which does help testing
and validating the program.

On the other hand, when the quality of the analysis is less than
100\%, it means that there exists at least one SCC where the
inferred termination condition is 0. Let us call such SCC's
\emph{failed SCC's}.
They are clearly identified, which may help the programmer.
Here are some reasons why cTI may fail:
potential non-termination,
poor numeric model,
non-existence of
a linear level mapping for a predicate with respect to the model,
inadequate norm.
Also, the analysis of the SCC's which depend on a failed SCC
is likely  to fail, but this does not
prevent cTI from analyzing other parts of the call graph.

\begin{table}
\renewcommand{\arraystretch}{1.2}
\centering
\caption{Running times for middle-sized programs.}
\label{table_results_middle}
\small
\begin{tabular}{||l|r||.|.|.|.|.|.||r||}
\hhline{~~|t:======:t|~}
\multicolumn{2}{l||}{} & \multicolumn{6}{c||}{analysis times (s)} \\
\hhline{|t:==:|------|:=:t|}
program & clauses
  & \multicolumn{1}{c|}{$M_P^{\cN}$}
  & \multicolumn{1}{c|}{$C_\mu$}
  & \multicolumn{1}{c|}{$\mu$}
  & \multicolumn{1}{c|}{$M_P^{\cB}$}
  & \multicolumn{1}{c|}{$\mathrm{TC}$}
  & \multicolumn{1}{c||}{total}
  & Q\% \\
\hhline{|:==::======::=:|}
\tt ann      & 177 & 0.17 & 0.48  & 0.08  & 0.17  & 0.06  & 1.00 & 49\%     \\ 
\tt bid      & 50 & 0.03  & 0.04  & 0.02  & 0.02  & 0.02  & 0.14    & 100\%    \\ 
\tt boyer    & 136 & 0.07  & 0.06  & 0.02  & 0.08  & 0.02  & 0.30     & 85\%     \\ 
\tt browse   & 30 & 0.05  & 0.12  & 0.03  & 0.04  & 0.01  & 0.26     & 60\%    \\ 
\tt credit   & 57 & 0.02  & 0.03  & 0.02  & 0.02 & 0.01  & 0.11    & 100\%    \\ 
\tt peephole & 134 & 0.18  & 0.56  & 0.03  & 0.20  & 0.06  & 1.08     & 94\%    \\ 
\tt plan     & 29 & 0.02  & 0.03  & 0.01  & 0.02  & 0.02  & 0.11     & 100\%    \\ 
\tt qplan    & 148 & 0.20  & 0.52  & 0.12  & 0.18  & 0.07  & 1.13      & 68\%  \\ 
\tt rdtok    & 55 & 0.13  & 0.39  & 0.03  & 0.07  & 0.02  & 0.65     & 44\%   \\ 
\tt read     & 88 & 0.26  & 1.00  & 0.04  & 0.31  & 0.08  & 1.72      & 52\%    \\ 
\tt warplan  & 101 & 0.10  & 0.25  & 0.01  & 0.08  & 0.02  & 0.49    & 33\%   \\ 
\hhline{|b:==::======::=:b|}
\multicolumn{2}{l||}{}
  & \multicolumn{1}{r|}{18\%}
  & \multicolumn{1}{r|}{50\%}
  & \multicolumn{1}{r|}{6\%}
  & \multicolumn{1}{r|}{17\%}
  & \multicolumn{1}{r|}{6\%}
  & \multicolumn{1}{r||}{100\%}  \\
\hhline{~~||------||}
\multicolumn{2}{l||}{} & \multicolumn{6}{c||}{average \% of time for each analysis phase} \\
\hhline{~~|b:======:b|}
\end{tabular}
\end{table}

\paragraph{Some larger programs.}
Finally, we have tested cTI on the following programs:
\begin{itemize}
\item {\tt chat} is a  parser  written by F.C.N.~Pereira and D.H.D.~Warren;
\item {\tt lptp} is an interactive theorem prover for Prolog written
by R.~St{\"a}rk \cite{Stark98};
\item {\tt pl2wam} is the compiler from Prolog to WAM of GNU-Prolog 1.1.2
developed by D.~Diaz \cite{DiazC00};
\item {\tt slice} is a multi-language interpreter developed
by R.~Bagnara and A.~Riaudo;
\item {\tt symbolic1} seems to be a simulator for a Prolog machine. We do not know the origin of this file.
\end{itemize}
The results of the analysis are given in
Table \ref{reslarger}. As explained in the previous section, we set up a timeout
of 2 seconds per SCC
for computing a CLP(\cN) model, the constraints defining level mappings,
a CLP(\cB) model, and the termination condition.
So we have a limit of 10 seconds
of CPU time per SCC.
The last but one column in Table~\ref{reslarger} summarizes
the number of timeouts for steps 2/3a/3b/4/5, respectively.

\begin{table}
\renewcommand{\arraystretch}{1.2}
\centering
\caption{Running times for larger programs.}
\label{reslarger}
\footnotesize
\begin{tabular}{||l|r||.|.|.|.|.|.||c||r||}
\hhline{~~|t:======:t|~}
\multicolumn{2}{l||}{} & \multicolumn{6}{c||}{analysis times (s)} \\
\hhline{|t:==:|------|:=:t:=:t|}
program & clauses
  & \multicolumn{1}{c|}{$M_P^{\cN}$}
  & \multicolumn{1}{c|}{$C_\mu$}
  & \multicolumn{1}{c|}{$\mu$}
  & \multicolumn{1}{c|}{$M_P^{\cB}$}
  & \multicolumn{1}{c|}{$\mathrm{TC}$}
  & \multicolumn{1}{c||}{total}
  & timeouts
  & Q\% \\
\hhline{|:==::======::=::=:|}
\tt chat    & 515 & 3.89   & 2.78   & 2.84   & 2.85  & 0.35  & 12.80   & 1/1/1/0/0 & 71\% \\ 
\tt lptp       & 1298 & 3.99   & 14.10  & 1.84  & 2.88  & 1.65  & 25.10   & 0/1/0/0/0 & 67\%  \\
\tt pl2wam     & 1190 & 2.12  & 2.37  & 0.99  & 2.00  & 1.29  & 9.22    & 0/0/0/0/0 & 64\% \\ 
\tt slice      & 952 & 2.20  & 10.46 & 0.13  & 2.08  & 0.93  & 16.20  & 0/3/0/0/0 & 55\% \\ 
\tt symbolic1  & 923 & 1.49  & 0.67  & 0.04  & 0.61  & 0.29  & 3.47 & 0/0/0/0/0 & 58\%  \\ 
\hhline{|b:==:b:======:b:=:b:=:b|}
\end{tabular}
\end{table}

\section{Related Work}
\label{sec:related-work}

The compiler of the Mercury programming language \cite{SomogyiHC96}
includes a termination checker, described in \cite{SpeirsSS97}.
The speed of the analyzer is quite impressive. We see two reasons for
this. First, the termination checker is written in Mercury itself.
Second, and most importantly, the analyzer takes high profit of the
mode informations that are part of the text of the program being
checked.
On the other hand, while the running times of cTI are bigger,
termination inference is a more general problem than termination
checking: in the worst case, an exponential number of termination
checks are needed to simulate termination inference.

TALP \cite{ArtsZ96} is an automatic tool that transforms a well-moded
logic program (see, e.g., \cite{Apt97}) into a term rewriting system
such that termination of the latter implies termination of the former.
The generated term rewriting system is then proved terminating by the
CiME tool (\url{http://cime.lri.fr/}).
The system seems quite powerful for this class of logic programs.

\cite{GenaimC01} made recently a link between backward analysis
\cite{KingL01} and termination analysis, which leads to termination
inference.
Although they used a completely different scheme for computing level
mappings, the results of the analysis on the programs described
in Tables~\ref{table_results_dds} and \ref{table_results_middle}
were rather similar, both in time and quality, to previous versions of cTI
that rely on the rational linear solver of SICStus
Prolog.
Thanks to the PPL, cTI is now significantly faster (speed-ups from
a factor of two to more than an order of magnitude have been observed).
The latest version of TerminWeb emphasizes termination analysis of typed
logic programs.

Termination of logic programming where numerical computations
are taken into account are studied in \cite{Serebrenik01,Serebrenik02}.
The authors present some advanced techniques for explicitly dealing with integers and
floating point numbers computations.

The \emph{size-change termination principle}
has been proposed in \cite{LeeJB-A01}
for deciding termination of first-order functional programs. The resulting
analysis is close to the TermiLog approach \cite{LindenstraussS97} and
the authors establish its intrinsic complexity.

Finally, we point out that the system Ciao-Prolog
\cite{BuenoCCHL-GP97}
adopts another approach for termination, based on complexity
analysis \cite{DebrayL-GHL94}.

\section{Conclusion}
\label{sec:conclusion}

We have presented cTI, the first bottom-up left-termination inference
tool for ISO-Prolog, and its experimental evaluation over standard
termination benchmarks as well as middle-sized and larger logic programs.
Running cTI on large programs shows that the approach scales up
satisfactorily.
We believe that, thanks to the Parma Polyhedra Library, cTI is today
the fastest and most robust termination inference tool for logic
programs.

When a SCC is too large, computations relying
on projection  may become too expensive.
So we have added for each computation which may be too costly
a timeout and if necessary we are able to return a value which
does not destroy the correctness of the analysis,
although the quality of the inference is obviously weaker.
It allows cTI to keep on analyzing the program.
As a side effect, the running time of cTI is  \emph{linear}
with respect to the number of SCC's in the call graph.

Finally, one can observe that the termination conditions computed
in Section \ref{sec:cti-overview} are actually \emph{optimal}
with respect to the language used for describing classes
of queries. Can one prove such properties automatically?
\cite{MesnardPN02} presents a first step in this direction.

\bigskip
\noindent
{\bfseries Acknowledgments.}
We would like to thank
Ulrich Neumerkel for numerous discussions we had on termination inference
and for the help he provided while debugging cTI.
Thanks also to the readers of a previous version of this paper
for their comments.

\bigskip
\noindent
{\bfseries Availability.}
cTI is distributed under the GNU General Public License.
The analyzer, together with the programs analyzed for benchmarking,
are available from cTI's web site: \url{http://www.cs.unipr.it/cTI}.

\hyphenation{ Ba-gna-ra Bie-li-ko-va Bruy-noo-ghe Common-Loops DeMich-iel
  Dober-kat Er-vier Fa-la-schi Fell-eisen Gam-ma Gem-Stone Glan-ville Gold-in
  Goos-sens Graph-Trace Grim-shaw Her-men-e-gil-do Hoeks-ma Hor-o-witz Kam-i-ko
  Kenn-e-dy Kess-ler Lisp-edit Lu-ba-chev-sky Nich-o-las Obern-dorf Ohsen-doth
  Par-log Para-sight Pega-Sys Pren-tice Pu-ru-sho-tha-man Ra-guid-eau Rich-ard
  Roe-ver Ros-en-krantz Ru-dolph SIG-OA SIG-PLAN SIG-SOFT SMALL-TALK Schee-vel
  Schlotz-hauer Schwartz-bach Sieg-fried Small-talk Spring-er Stroh-meier
  Thing-Lab Zhong-xiu Zac-ca-gni-ni Zaf-fa-nel-la Zo-lo
  }\newcommand{\noopsort}[1]{}

\end{document}